\documentclass[10pt,twocolumn,letterpaper]{article}

\usepackage{iccv}
\usepackage{times}
\usepackage{epsfig}
\usepackage{graphicx}
\usepackage{amsmath}
\usepackage{amssymb}
\usepackage{booktabs}

\DeclareMathOperator{\atantwo}{atan2}

\usepackage[pagebackref=true,breaklinks=true,letterpaper=true,colorlinks,bookmarks=false]{hyperref}

\iccvfinalcopy 

\def\iccvPaperID{1615} 

\begin{document}

\title{Fourier Space Losses for Efficient Perceptual Image Super-Resolution}

\author{Dario Fuoli$^1$ \qquad Luc Van Gool$^{1,2}$ \qquad Radu Timofte$^1$ \\
{\small $^1$ETH Zurich, Switzerland \qquad $^2$KU Leuven, Belgium}\\
{\tt\small \{dario.fuoli, vangool, radu.timofte\}@vision.ee.ethz.ch}}


\maketitle

\begin{abstract}
Many super-resolution (SR) models are optimized for high performance only and therefore lack efficiency due to large model complexity. As large models are often not practical in real-world applications, we investigate and propose novel loss functions, to enable SR with high perceptual quality from much more efficient models. The representative power for a given low-complexity generator network can only be fully leveraged by strong guidance towards the optimal set of parameters. We show that it is possible to improve the performance of a recently introduced efficient generator architecture solely with the application of our proposed loss functions.
In particular, we use a Fourier space supervision loss for improved restoration of missing high-frequency (HF) content from the ground truth image and design a discriminator architecture working directly in the Fourier domain to better match the target HF distribution. 
We show that our losses' direct emphasis on the frequencies in Fourier-space significantly boosts the perceptual image quality, while at the same time retaining high restoration quality in comparison to previously proposed loss functions for this task. The performance is further improved by utilizing a combination of spatial and frequency domain losses, as both representations provide complementary information during training. On top of that, the trained generator achieves comparable results with and is $2.4\times$ and $48\times$ faster than state-of-the-art perceptual SR methods RankSRGAN and SRFlow respectively.
\end{abstract}

\section{Introduction}
Super-resolution (SR) deals with the problem of reconstructing the high-frequency (HF) information from a low-resolution (LR) image $x\in\mathbb{R}^{H\times W\times C}$, which are inherently lost after downsampling the high-resolution (HR) image $y\in\mathbb{R}^{rH\times rW\times C}$ due to the lower Nyquist frequency in the LR space ($r$ denotes the scaling factor).
Recent single image SR (SISR) methods \cite{srcnn, vdsr, edsr, pyramid, back_proj, Hui-IMDN-2019} have shown remarkable success at reconstructing the missing HF details, with emphasis on accurate restoration of the frequency content in the ground truth frames. This is typically performed with supervised training, where the ground truth images $y$ are downsampled with a known kernel, \eg bicubic, to obtain the LR input images $x$.

While it may be desirable in some applications to restore the frequencies as close to the target as possible with minimal assumptions, the ill-posed problem limits the SR networks to generate higher frequency components, as the training promotes conservative estimates imposed by the pixel-wise supervision losses. This usually results in blurry images, which appear to be of lower quality than their respective HR counterparts.

\begin{figure}[t]
\begin{center}
   \includegraphics[width=1\linewidth]{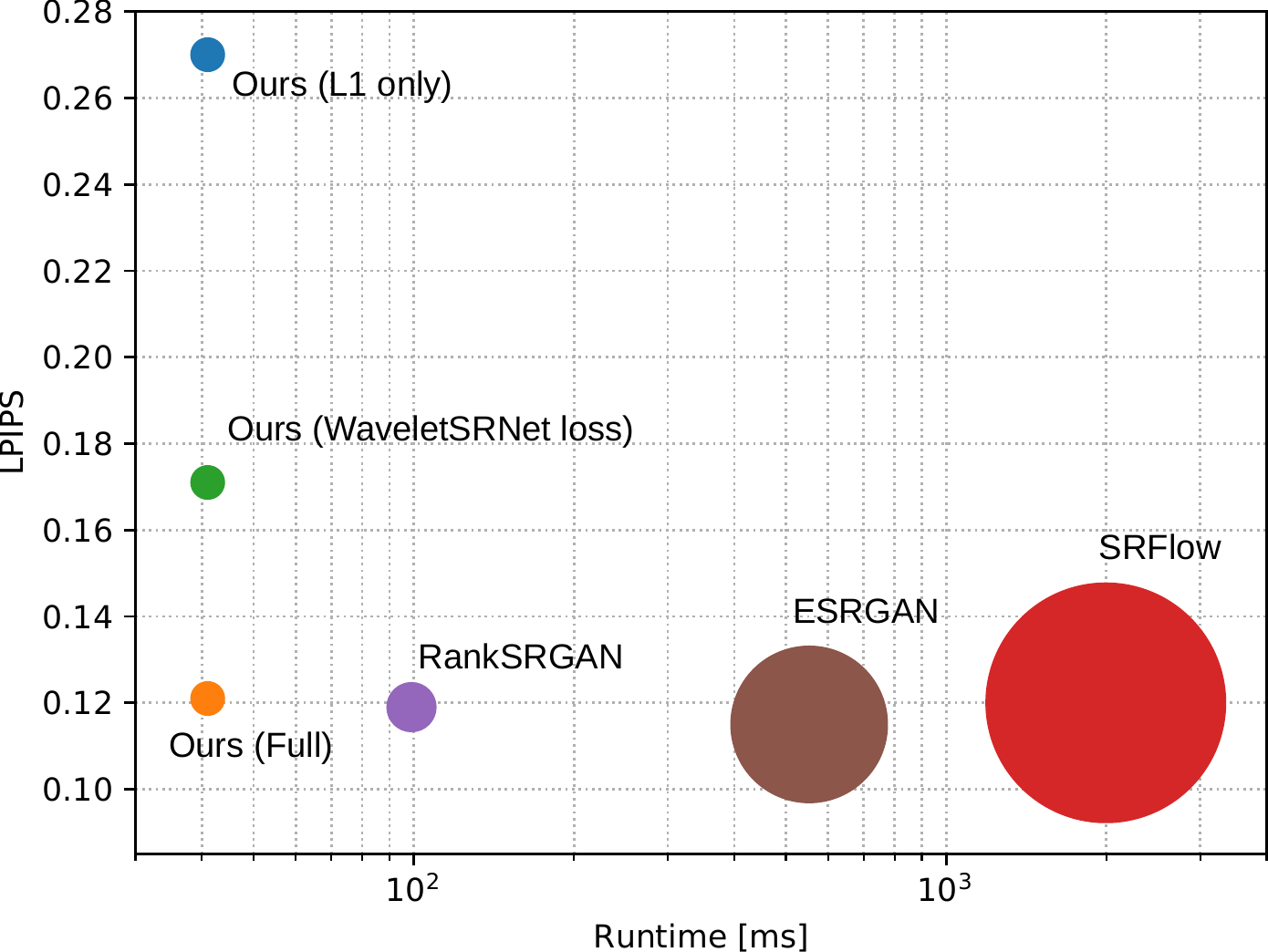}
\end{center}
   \caption{Runtime [ms] vs. perceptual quality (LPIPS)~\cite{zhang2018perceptual} comparison with state-of-the-art methods. The disk area is proportional to the number of parameters. We achieve the fastest runtimes with comparable perceptual quality to much larger networks.}
\label{fig:eye_catcher}
\end{figure}

This issue has been addressed in the literature \cite{srgan, wang2018esrgan} by employing different losses, that are designed to promote the higher frequencies for perceptually more pleasing images. These supervised objectives are often used in combination with generative adversarial networks~\cite{gan} (GAN) for additional distribution learning of the HF space. Conditional GAN-based learning enables the generation of plausible high frequencies without the need for ground truth accuracy. A lot of research has been devoted to design such perceptual losses and to find suitable combinations for pleasing results. 

Today, more and more deep learning based algorithms are implemented on smartphones, which requires low-complexity networks for fast inference and inexpensive deployment. Therefore, the design focus is slowly shifting from high-quality, high-performance methods with high-complexity networks to more efficient enhancers, which upscale much faster and require less resources. 
In contrast to empowering a deep neural network's performance by simply increasing it's complexity, which is generally straight-forward, finding an efficient network with high-performance is a much harder challenge. Searching for effective low-complexity networks with high performance, that are on par with state-of-the-art methods, is the ultimate challenge in network design. 

Three main ingredients are necessary in order to maximize performance and efficiency of deep neural networks. First, the best architecture design for the task has to be determined. Usually, this task is performed manually by experts. In addition to handcrafted designs, neural architecture search algorithms \cite{Gong_2019_ICCV_autogan, autogandistiller} have recently been proposed to automate this task. Second, the design of the optimal loss function is imperative to fully leverage a network's performance. Third, the amount and quality of data plays a key role to maximize performance.  A large portion of existing literature in SR deals with the first point. We regard the solution to the third point as straight-forward, as data can be collected efficiently for most applications. In this paper we propose a solution to the second point and try to maximize the performance of a recently proposed efficient low-complexity network \cite{Hui-IMDN-2019, ZhangICCVW2019} for perceptual SR, solely by the application of our proposed loss functions.


The design of perceptual losses predominantly focuses on the spatial domain~\cite{wang2018esrgan, srgan}. However, SR is tightly coupled to the frequency domain, as only high frequencies are removed during the downsampling process. We leverage this fact and propose novel loss functions in Fourier space by calculating the frequency components with the fast Fourier transform (FFT) for direct emphasis on the frequency content. 
We propose a supervision loss in direct reference to the ground truth directly in Fourier domain for accurate reconstruction. Additionally, we propose a discriminator architecture to learn the HF distribution in an adversarial training setup working directly in Fourier space. To the best of our knowledge we are the first to apply a GAN loss directly on Fourier coefficients in SR.
Our ablation study shows clear benefits over spatial losses for the task of perceptual SR.
Also, employing a loss in Fourier space introduces global guidance as opposed to pixel-wise evaluation due to the nature of the Fourier transform. In order to leverage both global and local guidance, we also add the corresponding spatial supervision and GAN losses. Together with an additional perceptual loss, this outperforms all other configurations in our ablation study.
In addition to the advantage of our proposed losses over existing ones, we compare our trained efficient generator with high-performance state-of-the-art methods. It shows, that our losses can substantially increase the performance of a low-complexity generator to even compete with much larger networks.

\section{Related Work}
SR is a popular topic and a series of competitions are conducted by \cite{Timofte_2017_CVPR_Workshops, Agustsson_2017_CVPR_Workshops, Blau_2018_ECCV_Workshops, ZhangICCVW2019, aim2020_efficient_SR, Zhang_2020_CVPR_Workshops, Lugmayr_2020_CVPR_Workshops} which provide a broad overview of research and development over recent years in this area.

\noindent\textbf{Restoration}
Learning based approaches have shown to be highly effective so solve the problem of SR and are therefore predominantly used in research. SRCNN~\cite{srcnn} is one of the first convolutional neural network (CNN) based methods to surpass non-CNN SR algorithms, VDSR~\cite{vdsr} is an improved version which adopts a deeper network for improved performance. Further concepts and improvements are explored~\cite{srgan, edsr, pyramid, back_proj, Hui-IMDN-2019} with the aim of reconstructing the missing details in a LR image as close to the ground truth as possible.

\noindent\textbf{Perceptual SR}
Since even the best of the aforementioned methods tend to produce blurry images, another family of methods~\cite{srgan, wang2018esrgan, Zhang_2019_ICCV} tries to further improve the perceptual image quality by sacrificing restoration quality for increased generation of HF content~\cite{Blau_2018_ECCV_Workshops}. 
For that matter, SRGAN~\cite{srgan} proposes the application of a generative adversarial network (GAN)~\cite{gan} to better model the HF distribution in an image. The authors also propose a perceptual loss, based on features of VGG~\cite{Simonyan2015vgg}, which significantly boosts the perceptual quality. ESRGAN~\cite{wang2018esrgan} extends this concept by adopting an improved GAN-loss formulation~\cite{jolicoeur2018relativistic} and a stronger generator architecture. 
RankSRGAN~\cite{Zhang_2019_ICCV} is another approach to achieve improved perceptual image quality. It uses a ranker to enable gradient based training with non-differentiable handcrafted no-reference image quality metrics. First, a dataset with pairs of images and their calculated quality score is prepared, then a ranker is trained to relatively rank two images in a differentiable manner. The learned differentiable ranker is then used in a gradient based adversarial training setup.
More recently, SRFlow~\cite{lugmayr2020srflow} uses normalizing flows~\cite{normalizing_flows} for perceptual image SR. The method explicitly models the ambiguity in HR space and is trained by maximimum likelihood with the use of a network that is invertible by design. 

\noindent\textbf{Frequency-based SR}
Since SR is the problem of restoring frequency components, several works \cite{huang2017wavelet, fritsche2019_fsep, Guo_2017_CVPR_Workshops, jiang2020focal} propose to model the problem closer to frequency space in various configurations. WaveletSRNet~\cite{huang2017wavelet} uses wavelets to decompose the LR image by the Haar transform and generates the missing HF wavelet coefficients instead of HR images directly. Additionally, the losses are optimized for perceptual image quality by weighing the wavelet coefficients by some heuristic, in order to balance the importance of different sub-bands. DWSR~\cite{Guo_2017_CVPR_Workshops} uses a similar approach without a weighting scheme and uses only four sub-bands, without explicit perceptual components. The loss in \cite{huang2017wavelet} is composed of more sub-bands, but it does not fully decompose the image as we do by applying the Fourier transform.
A more recent work \cite{jiang2020focal} proposes a supervision loss in Fourier space as additional loss for generative tasks. However, this work uses a different loss formulation, \ie it computes the differences directly between the complex components without transformation into amplitude and phase. On top of that, to the best of our knowledge, we are the first to also employ a GAN loss directly in Fourier space.



\section{Proposed Method}
\label{sec:method}

\begin{figure*}[t]
\centering
   \includegraphics[width=0.9\linewidth]{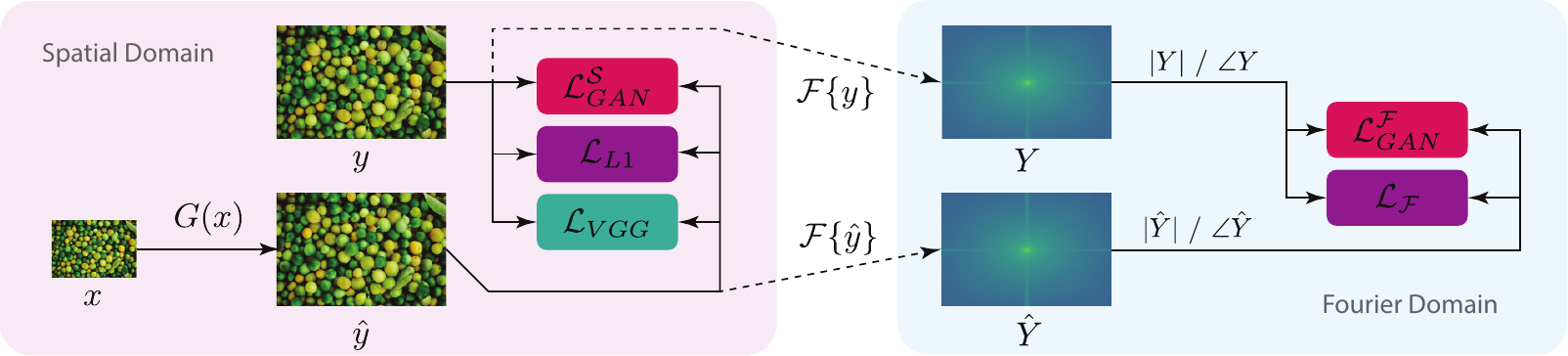}
   \caption{Overview of the proposed method. We employ losses in both spatial and Fourier domain to strengthen the training signal.}
\label{fig:overview}
\end{figure*}

The task of image SR, is to increase the resolution of an image $x\in\mathbb{R}^{H\times W\times C}$ from the LR domain $\mathcal{X}$ to the corresponding image $y\in\mathbb{R}^{rH\times rW\times C}$ in HR domain $\mathcal{Y}$ with factor $r$. According to Nyquist–Shannon's sampling theorem, the missing HF content above the Nyquist-frequency $n_c$ must be recovered in order to get an image $y$ from the target HR domain $\mathcal{Y}$.
In contrast to the representation of an image in spatial domain, these missing frequencies can be clearly separated in Fourier domain. We therefore propose two losses in the frequency domain, to directly emphasize the training on the relevant frequencies. Additionally, the frequency components provide global guidance during training due to the nature of the Fourier transform. 

\subsection{Generator}
Our aim is to reduce the computational complexity of the generator network for faster runtimes, while retaining the representational power for SR as high as possible. Therefore, the design of more effective losses is imperative. Improving the loss design can yield stronger gradient signals which better guide the generator during the training process. In order to test the effectiveness of our proposed losses, we use a lightweight model based on the IMDN network~\cite{Hui-IMDN-2019} from the same authors. This is the winner of the ``AIM 2019 Challenge on Constrained SR''~\cite{ZhangICCVW2019}. The network is used as an example of an efficient generator architecture to showcase the power of our loss designs against typical existing losses.
The network consists of repeated information multi-distillation blocks (IMDBs), that are designed to effectively integrate information from the LR-space towards the HR-space. The whole processing is conducted in LR-space for efficiency reasons. Only in the last processing step, the refined HR image is upsampled with a standard shuffling block~\cite{shuffling}.
generator $G$ super-resolves a LR image $x\in\mathbb{R}^{H\times W\times C}$ into a HR image $\hat{y}=G(x)\in\mathbb{R}^{rH\times rW\times C}$. 

\subsection{Fourier Transform and SR}
\label{sec:fftandsr}
The Fourier transform is widely used to analyze the frequency content in signals. It can also be applied to multi-dimensional signals such as images, where the spatial variations of pixel-intensities have a unique representation in the frequency domain. The discrete Fourier transform (DFT) decomposes an image $x\in\mathbb{R}^{H\times W\times C}$ from the spatial domain into the Fourier domain. The Fourier space is spanned by complex orthonormal basis functions, where the complex frequency components $X\in\mathbb{C}^{U\times V\times C}$ characterize the image. 

\begin{equation}
    \mathcal{F}\{x\}_{u,v} = X_{u,v} = \frac{1}{\sqrt{HW}}\sum_{h=0}^{H-1}\sum_{w=0}^{W-1} x_{h,w}  e^{-i2\pi\left(u\frac{h}{H} + v\frac{w}{W} \right)} 
\end{equation}

Since images are composed of multiple color channels, we calculate the Fourier transform for each channel separately and perform the transform per channel. The explicit notation of channels is omitted in our formulas.
Each complex component $X_{u,v}$ can be represented by amplitude $|\mathcal{F}\{x\}_{u,v}|$ and phase $\angle \mathcal{F}\{x\}_{u,v}$, which provides a more intuitive analysis of the frequency content.

\begin{equation}
    |\mathcal{F}\{x\}_{u,v}| = |X_{u,v}|= \sqrt{\mathcal{R}\{X_{u,v}\}^2 + \mathcal{I}\{X_{u,v}\}^2}
\end{equation}
\begin{equation}
    \angle \mathcal{F}\{x\}_{u,v} = \angle X_{u,v} = \atantwo (\mathcal{I}\{X_{u,v}\}, \mathcal{R}\{X_{u,v}\})
\end{equation}

Due to symmetry in the Fourier space (Hermitian symmetry) for real valued signals $x$, we can omit redundant spectral components and only treat half of $X$, and still retain the full information in $x$. 

\begin{equation}
\label{eq:hermitian}
   \mathcal{F}\{x\}_{u,v} = \overline{\mathcal{F}\{x\}_{-u,-v}}
\end{equation}

Thus, processing can be significantly reduced by neglecting redundant components when working in the Fourier domain of real-valued signals like images.
Note, despite discarding the redundant values, the total number of values in the spatial and Fourier domain remains the same since the components in Fourier space are composed of real and imaginary part (or amplitude and phase).

Since the Fourier transformation assumes an infinite signal in the transformation dimensions, finite signals like images should be preprocessed to avoid edge induced artifacts.
We avoid such artifacts by applying a Hann window, which suppresses the signals' amplitude towards the edges in order to smooth out the transitions. Afterwards, the image is transformed with a more accurate representation of the frequency spectrum.

As SR is the task of reconstructing the missing HF content from a downscaled image, a reduction in the sampling rate leads to a lower Nyquist-frequency $n_c$ in the LR-space, which constitutes a hard limit in the representation capability of high frequencies above said frequency. Therefore, SR deals with the problem of generating these missing frequencies, which can be seen as the extrapolation from low to high frequencies. Contrary to the representation of an image in the spatial domain, these frequencies can be clearly separated in the frequency space in order to directly emphasize the important image features for SR. Additionally, the Fourier components provide global information about the image as opposed to local information represented by pixel values in the spatial domain.
We leverage these properties to design new losses for efficient perceptual SR training.

In contrast to the Fourier transform, wavelet-transforms balance local- and global frequency precision in an image by dividing the frequency content into different sub-bands. This property is useful for many practical applications where this information is needed. However, we are not forced to find a balance. For application in our losses, we can both leverage the global frequency content with maximal precision represented by one component for each frequency in the signal and get precise local guidance through the spatial representation of the image.

\subsection{Supervision Losses}

For perceptual SR, predominantly spatial domain based losses, spatial feature losses, or frequency-band separation strategies in the spatial domain, \eg separation by wavelet decomposition or filtering, are proposed~\cite{wang2018esrgan, fritsche2019_fsep}. Presumably, because most existing architectures are based on convolutions that expect spatial invariance in the input and also due to easy handling of variable image sizes of convolutional networks. Popular choices for supervision losses, \ie with reference to a ground truth, are pixel-based losses L1/L2 and feature based VGG-loss~\cite{Simonyan2015vgg, srgan}. As proposed in \cite{wang2018esrgan} and for direct comparison, we investigate L1~\eqref{eq:l1} and VGG-loss~\eqref{eq:vgg}.

\begin{equation}
\label{eq:l1}
    \mathcal{L}_{L1} = \frac{1}{HW}\sum_{h=0}^{H-1}\sum_{w=0}^{W-1}\left|\left|\hat{y}_{h,w}-y_{h,w}\right|\right|_1
\end{equation}

\begin{equation}
\label{eq:vgg}
    \mathcal{L}_{VGG} = \frac{1}{IJ}\sum_{i=0}^{I-1}\sum_{j=0}^{J-1}\left|\left|N_{VGG}^{54}(\hat{y})_{i,j}-N_{VGG}^{54}(y)_{i,j}\right|\right|_1
\end{equation}

Following the setting in \cite{wang2018esrgan} we calculate a VGG-loss using the pre-trained 19-layer VGG network. In particular, the L1-norm between features $N_{VGG}^{54}(\cdot)$ (54 indicates 4th convolution before the 5th pooling layer) from generator output $\hat{y}=G(x)$ and the target $y$ constitutes the VGG-loss. 

In addition to these spatial domain losses, we propose a Fourier space loss $\mathcal{L}_\mathcal{F}$ for supervision from the ground truth frequency spectrum during training. First, ground truth $y$ and generated image $\hat{y}$ are pre-processed with a Hann window, as described in Section~\ref{sec:fftandsr}. Afterwards, both images are transformed into Fourier space by applying the fast Fourier transform (FFT), where we calculate amplitude and phase of all frequency components. The L1-norms of amplitude difference $\mathcal{L}_{\mathcal{F}, |\cdot|}$ and phase angle difference $\mathcal{L}_{\mathcal{F}, \angle}$ between output image and target are averaged to produce the total frequency loss $\mathcal{L}_\mathcal{F}$. 
Note, since 50\% of all frequency components are redundant, the summation for $u$ is performed up to $U/2-1$ only, without affecting the loss due to Eq.~\eqref{eq:hermitian}.

\begin{equation}
\label{eq:}
    \mathcal{L}_{\mathcal{F}, |\cdot|} = \frac{2}{UV}\sum_{u=0}^{U/2-1}\sum_{v=0}^{V-1}\left|\left||\hat{Y}|_{u,v}-|Y|_{u,v}\right|\right|_1
\end{equation}

\begin{equation}
    \mathcal{L}_{\mathcal{F}, \angle} = \frac{2}{UV}\sum_{u=0}^{U/2-1}\sum_{v=0}^{V-1}\left|\left|min(\angle \hat{Y}_{u,v}-\angle Y_{u,v})\right|\right|_1
\end{equation}

\begin{equation}
    \mathcal{L}_\mathcal{F} = \frac{1}{2}\mathcal{L}_{\mathcal{F}, |\cdot|} + \frac{1}{2}\mathcal{L}_{\mathcal{F},\angle}
\end{equation}

Theoretical benefits of applying a supervision loss in Fourier domain are two-fold. (1) The direct emphasis, especially on the missing HF components, promotes generation in these important areas as opposed to spatial losses (L1/L2), which are known to produce blurry images. (2) Due to the nature of the Fourier transform, which computes the frequency content with highest precision in trade-off for spatial precision, the loss is directly calculated on the global frequency components and therefore provides global guidance during training in contrast to local pixel-based losses in spatial domain.

In theory, it would be sufficient to calculate the loss between the HF components only, due to Nyquist–Shannon’s sampling theorem (see Sec.~\ref{sec:method}). However, because of resampling and (imperfect) anti-aliasing filtering during the downscaling process, there will be mismatches between the low-frequency components as well. Applying the loss also on these frequencies, improves the restoration of low-frequency content in $\hat{y}$. 

In contrast to other frequency-based losses, proposed in the literature, we directly apply the losses in Fourier space, and do not tune our losses according to some heuristic, as in \cite{huang2017wavelet}.






\subsubsection{GAN Losses}
\label{sec:gan_losses}

\begin{figure}[t]
\begin{center}
   \includegraphics[width=0.9\linewidth]{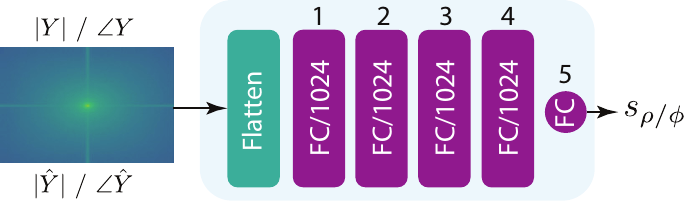}
\end{center}
   \caption{Proposed Fourier GAN architecture. We process the Fourier components of $y$ and $\hat{y}$ with a fully connected network to predict real $s_\rho$ and fake $s_\phi$ scores.}
\label{fig:FourierGAN}
\end{figure}

In order to further boost the perceptual quality we employ a GAN training scheme with two types of GAN-architectures, applied in spatial and Fourier domain. Learning the mapping from LR to HR directly from the ground truth severely limits the generation of images with high perceptual quality. Minimizing the risk towards a single realisation represented by the ground truth is too strict because the problem is ill-posed. A GAN training strategy relaxes the loss formulation by allowing plausible HR reconstructions resembling images from the target distribution.

We use the discriminator from \cite{wang2018esrgan} for our spatial GAN loss $\mathcal{L}^{\mathcal{S}}_{GAN}$. Additionally, we design a discriminator working directly in Fourier domain for our proposed frequency domain GAN-loss $\mathcal{L}^{\mathcal{F}}_{GAN}$. After the transformation of an image into Fourier space, the spatial invariance assumption is no longer valid. Therefore, the application of a convolutional architecture will not be optimal for this task. Thus, we apply a fully connected discriminator network for adversarial guidance in Fourier space, see Fig.~\ref{fig:FourierGAN}. Again, generated image $\hat{y}$ and ground truth $y$ are transformed into frequency components represented by amplitude and phase in Fourier space after the application of a Hann window.  

Both adversarial losses are evaluated by a relativistic GAN formulation~\cite{jolicoeur2018relativistic}, which showed improved performance in SR over the standard GAN formulation in \cite{wang2018esrgan}.
The discriminator's real and fake logits $s_\rho = D(y), s_\phi = D(\hat{y})=D(G(x))$ are processed with the relativistic transformation by averaging the logits over the batch dimension $b$, Eq.~\eqref{eq:rel_sum}, and subtracting them from the original logits, Eq.~\eqref{eq:rel_trans}. The transformed real and fake scores $\rho, \phi$ are then evaluated with the standard sigmoid cross-entropy GAN-objective in \eqref{eq:gan_objective}.


\begin{equation}
    \label{eq:rel_sum}
    \overline{s}_\rho=\frac{1}{B}\sum_{b}^{B}s_\rho(b), \quad
    \overline{s}_\phi=\frac{1}{B}\sum_{b}^{B}s_\phi(b)
\end{equation}

\begin{equation}
\label{eq:rel_trans}
\begin{aligned}
\rho = & D(y)-\overline{s}_\phi, \quad \phi = D(\hat{y}) - \overline{s}_\rho
 \end{aligned}
\end{equation}

\begin{equation}
\label{eq:gan_objective}
\begin{aligned}
\mathcal{L}^{G}_{GAN} = & -\mathbb{E}_{x,y}\left[\log(\sigma\left(\phi\right)) + \log(1 - \sigma\left(\rho\right))\right] \\
\mathcal{L}^{D}_{GAN} = & -\mathbb{E}_{x,y}\left[\log(\sigma\left(\rho\right)) + \log(1 - \sigma\left(\phi\right))\right]
\end{aligned}
\end{equation}


\subsection{Training Setup}

The complete training setup \eqref{eq:gan_objective_complete} consists of two supervision losses and two GAN-losses in both spatial and Fourier domain and an additional VGG-loss. These loss components are weighted with factors $\alpha, \beta, \gamma$ and minimized with Adam~\cite{adam} optimizer in alternating steps.

\begin{equation}
\label{eq:gan_objective_complete}
    \begin{aligned}
    \min_{G}\text{ } & \alpha\left(\frac{\mathcal{L}^{G,\mathcal{S}}_{GAN} +  \mathcal{L}^{G,\mathcal{F}}_{GAN}}{2}\right) + \beta\left(\frac{\mathcal{L}_{L1} + \mathcal{L}_{\mathcal{F}}}{2}\right) + \gamma\mathcal{L}_{VGG} \\ 
    \min_{D}\text{ } & \alpha\left(\frac{\mathcal{L}^{D,\mathcal{S}}_{GAN} +  \mathcal{L}^{D,\mathcal{F}}_{GAN}}{2}\right)
    \end{aligned}
\end{equation}

\section{Experiments and Results}

\begin{table*}
\begin{center}
\begin{tabular}{clccccccccc}
\toprule
Configuration & Generator & $\mathcal{L}_{L1}$ & $\mathcal{L}_{F}$ & $\mathcal{L}^{\mathcal{S}}_{GAN}$ & $\mathcal{L}^{\mathcal{F}}_{GAN}$ & $\mathcal{L}_{VGG}$ &  $\uparrow$PSNR & $\uparrow$SSIM & $\downarrow$LPIPS
& $\downarrow$FID
\\
\midrule

1 & IMDN \cite{Hui-IMDN-2019} & \checkmark &  & & && \textcolor{red}{30.56} & \textcolor{red}{0.837} & 0.270 & 22.91\\

2 & IMDN \cite{Hui-IMDN-2019} & & \checkmark &&&& 29.53 & 0.811 & 0.189 & 16.98 \\

3 & IMDN \cite{Hui-IMDN-2019} & \checkmark & \checkmark &&&& \textcolor{blue}{30.32} & \textcolor{blue}{0.834} & 0.266 & 21.96 \\

4 & IMDN \cite{Hui-IMDN-2019}& \checkmark & & \checkmark & & \checkmark & 27.94  & 0.751 & 0.131  & 17.07 \\

5 & IMDN \cite{Hui-IMDN-2019} &  & \checkmark &  & \checkmark & \checkmark & 29.06 & 0.796 &0.129& 17.17 \\

6 & IMDN \cite{Hui-IMDN-2019}& \checkmark & \checkmark & \checkmark &  & \checkmark & 27.96
 & 0.762 & \textcolor{blue}{0.127} & \textcolor{blue}{16.94}\\

7 & IMDN \cite{Hui-IMDN-2019}& \checkmark & \checkmark &  & \checkmark & \checkmark & 29.13 & 0.794 & \textcolor{blue}{0.127}
& 17.90\\

8 & IMDN \cite{Hui-IMDN-2019}& \checkmark & \checkmark & \checkmark & \checkmark & \checkmark & 28.42 & 0.776 & \textcolor{red}{0.124} & \textcolor{red}{15.88} \\
\midrule
9 & ESRGAN \cite{wang2018esrgan} & \checkmark & \checkmark & \checkmark & \checkmark & \checkmark & 28.63 & 0.780 & 0.113 & 14.80 \\
10 & ESRGAN \cite{wang2018esrgan} & \checkmark &  & \checkmark &  & \checkmark & 28.19 & 0.769 & 0.115 & 15.37 \\

\bottomrule
\end{tabular}
\end{center}
\caption{Ablation study results. We compare different configurations of loss functions. 
We calculate restoration metrics PSNR and SSIM, perceptual metric LPIPS~\cite{zhang2018perceptual} and distributional similarity by FID~\cite{Heusel2017fid}. The metrics are calculated on the DIV2K validation dataset.}
\label{tab:ablation}
\end{table*}

All settings 
are trained on the DF2K dataset with a scaling factor of $r=4$. DF2K is a combination of DIV2K~\cite{Agustsson_2017_CVPR_Workshops} and Flickr2K~\cite{Timofte_2017_CVPR_Workshops}. Training pairs consist of paired crops of size $64\times64$ and $256\times256$ from LR and HR respectively. We evaluate all experiments on the DIV2K validation set, the standard benchmark for HR image SR. Additionally, we provide results on Urban100~\cite{Huang-CVPR-2015_urban100}.

We calculate restoration metrics PSNR and SSIM, perceptual metric LPIPS~\cite{zhang2018perceptual} and distributional similarity by FID~\cite{Heusel2017fid, obukhov2020torchfidelity}. We deliberately refrain from using no-reference metrics, since we want to learn the image quality from the target domain $\mathcal{Y}$, which is different to learning for a no-reference metric, as these handcrafted metrics do not necessarily correlate with the properties of the target image distribution.


\subsection{Ablation}

We conduct an ablation study with different loss configurations to show the effectiveness of our proposed Fourier domain losses, see Tab.~\ref{tab:ablation}. The generator is initialized with pretrained weights (L2) in all configurations and trained on DF2K for 500k iterations with a constant learning rate $l=10^{-5}$ and a batch size of $B=16$. We do not use a learning rate scheduler for stability reasons and fairness due to the heterogeneous combinations of different loss types. The training parameters are set to $\alpha=0.005$, $\beta=0.01$ and $\gamma=1$ as proposed in state-of-the-art  method ESRGAN~\cite{wang2018esrgan}. The averaging by factor 2 in \eqref{eq:gan_objective_complete} is removed whenever a single loss is employed per parameters $\alpha$ or $\beta$, to keep the balance between supervision and GAN-losses in all configurations. Additionally, we refine the pretrained generator from ESRGAN with our additional losses in the same setting with $B=8$.

A comparison between configuration 1 and 2 clearly shows the effectiveness of our proposed Fourier domain supervision loss $\mathcal{L_F}$ for perceptual quality enhancement. Calculating the losses with our proposed formulation significantly improves the perceptual image quality in trade-off with restoration quality~\cite{Blau_2018_ECCV_Workshops}, which is reflected by the large improvement of LPIPS (-0.081) and FID (-5.93).

Configuration 4 represents the loss formulation from ESRGAN~\cite{wang2018esrgan}, these spatial losses are exchanged by our proposed Fourier domain losses $\mathcal{L_F}$ and $\mathcal{L}^{\mathcal{F}}_{GAN}$ in configuration 5. The perceptual quality remains comparable between the two configurations. However, the restoration quality is significantly higher compared to the ESRGAN losses by a large margin, reflected by a gain in PSNR and SSIM of +1.12dB and +0.045 respectively, which already shows the superiority of our proposed Fourier domain losses over the corresponding spatial losses employed in ESRGAN.

Configuration 8 shows the effectiveness of our proposed Fourier domain losses in combination with spatial losses. It achieves the best LPIPS and FID scores of all configurations and clearly outperforms the losses of ESRGAN in configuration 4 in all metrics. Simultaneous application of losses in both spatial and frequency domain leverages complementary information from each image representation to significantly improve overall guidance during training. Configuration 9 shows the combination of ESRGAN generator with our proposed full combination of Fourier domain and spatial losses. We note the improvement (PSNR +0.44dB, FID -0.57) brought by our Fourier domain losses over the original ESRGAN in configuration 10.





\subsection{Comparison with State-of-the-art}

In addition to the effectiveness of our losses for performance in perceptual SR in our ablation study, we show that we can also compete with state-of-the-art methods in terms of performance with a more efficient generator network, due to our better losses. We tweak the loss weights towards higher perceptual quality in trade-off with restoration quality and set them to $\alpha=0.0025,\beta=0.005,\gamma=1$ for our model in Tab.~\ref{tab:results} and Tab.~\ref{tab:urban}. Note, the proposal of our losses is to showcase the improved training performance which enables to train high-performance low-complexity generators, not necessarily to achieve state of the art performance.
Despite the low complexity of $G$ in our setting, we are able to compete with image quality of the best state-of-the-art methods, with a substantial reduction of runtime. 

\textbf{ESRGAN} uses a combination of L1, VGG and GAN loss and propose an improved generator architecture derived from SRGAN~\cite{srgan}.
\textbf{RankSRGAN}~\cite{Zhang_2019_ICCV} proposes a method to use non-differentiable handcrafted image quality metrics (Ma~\cite{Ma_Metric_2017}, NIQE~\cite{NIQE} and PI~\cite{Blau_2018_ECCV_Workshops}) for training in a GAN-based setup. The generator network in RankSRGAN is SRGAN~\cite{srgan}. \textbf{SRFlow}~\cite{lugmayr2020srflow} is a recently proposed method, which uses normalizing flows~\cite{normalizing_flows} for perceptual image SR. The concept of normalizing flows provides an alternative to GAN-based learning by modeling the ill-posed problem explicitly as a stochastic process.
We also compare our loss formulation to recently proposed losses using the wavelet transformation, see Sec.~\ref{sec:fftandsr}. Division into subbands with wavelet transform is used by \textbf{WaveletSRNet}~\cite{huang2017wavelet} and DWSR~\cite{Guo_2017_CVPR_Workshops}, which both use the Haar transform. We compare our method to the losses in WaveletSRNet which uses a finer division and a more sophisticated loss formulation than DWSR. For this purpose, we train the efficient generator backbone $G$ with the proposed losses in WaveletSRNet for direct comparison. 

\begin{figure*}[t]
\begin{center}
   \includegraphics[width=1\linewidth]{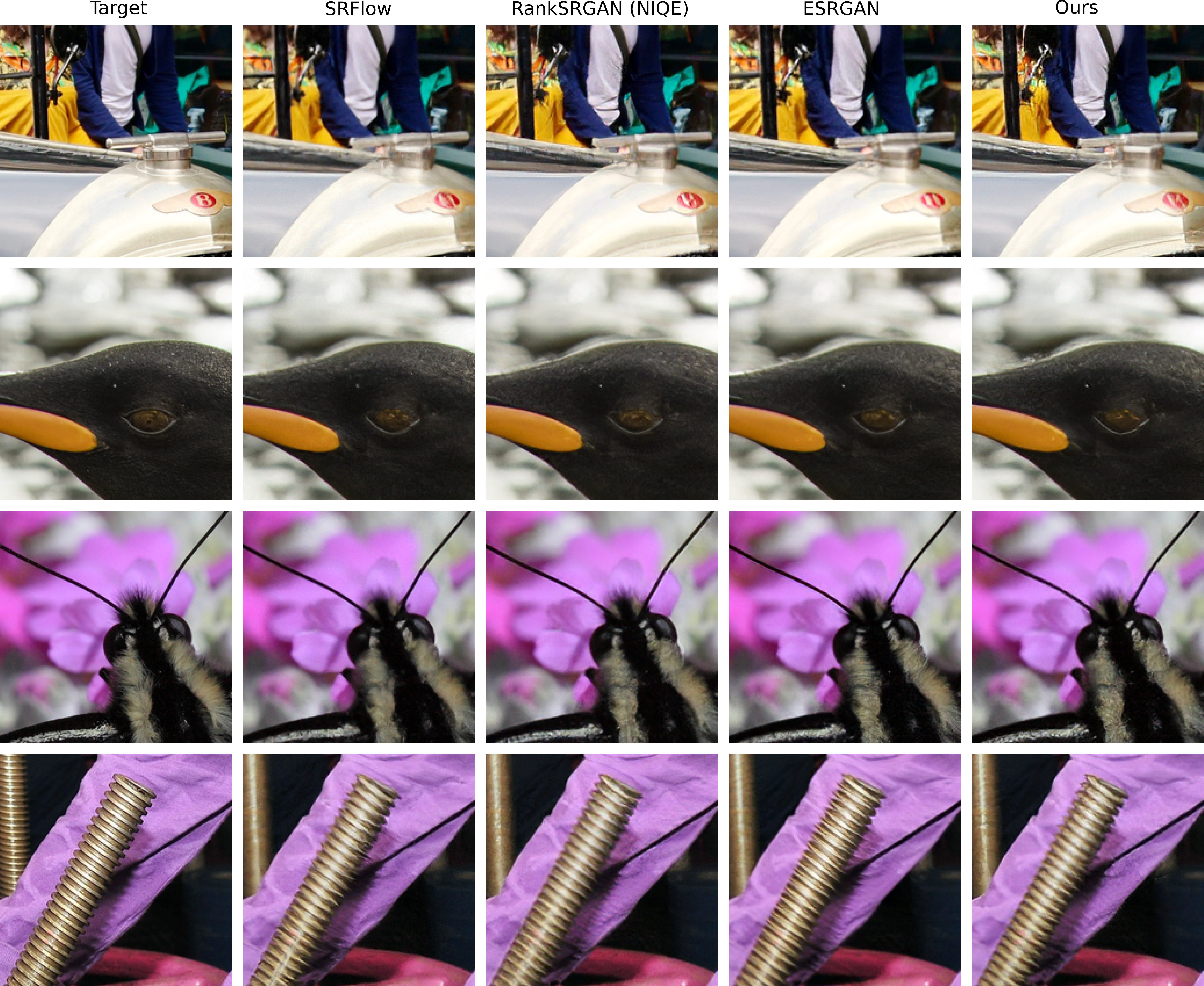}
\end{center}
   \caption{Visual examples on DIV2K validation images. }
\label{fig:visual_results}
\end{figure*}

For all other methods we use the pretrained models provided by the authors, as all of them are trained on DF2K. Additionally, we provide the results for standard bicubic upsampling as a baseline. To quantify the efficiency and model complexity, we compute runtimes and number of parameters
at inference on a NVIDIA TITAN RTX and an Intel i7 CPU (6 cores). 
We also provide visual examples in Fig.~\ref{fig:visual_results} which support our quantitative evaluation.


\subsubsection{Discussion}


The superiority of our losses compared to ESRGAN's losses is already shown in the ablation study in Tab.~\ref{tab:ablation}. On top of that, we can even compete with ESRGAN's high-complexity generator, which achieves slightly better LPIPS and FID values, but lower PSNR and SSIM scores with a substantially slower inference time by a factor of over $13\times$ on GPU.

Our losses significantly surpass all three RankSRGAN models in both restoration metrics PSNR/SSIM and even achieve the highest FID score. Only the NIQE and PI optimized models have slightly higher LPIPS scores, which however comes with a $2.4\times$ higher runtime on GPU. Note, this is a substantial difference in complexity, \eg this equates to reducing the number of layers in a network by a factor of $2.4$. In comparison to the ranker approach, our loss formulation does not depend on the difficult design of a meaningful handcrafted quality metric, which manifests an upper bound on the achievable quality. We also do not require the expensive setup of the ranker, however we achieve stronger guidance by direct emphasis on the frequency content without an additional explicit concept of perceptual quality. 

SRFlow~\cite{lugmayr2020srflow} is by far the most expensive method with a large number of parameters and slow inference speeds of 1.995s and 55.33s on GPU and CPU respectively, yet does not outperform the performance of other methods, with the exception of PSNR and SSIM. Our highly efficient method is on par with SRFlow with comparable  perceptual metrics. Our solution has better FID score (+0.41) but slightly lower LPIPS score (-0.001). However, there is an enormous difference in inference speed, \eg SRFLow is 48 times slower on GPU than our method, which clearly highlights the superiority of our proposed losses.  



We train $G$ with WaveletSRNet's losses from scratch with a learning rate of $l=10^{-5}$ for 500k iterations with a batch size of $B=16$. Further, we finetune $G$ with a lower learning rate of $l=10^{-6}$ for another 250k iterations. We clearly outperform WaveletSRNet's loss formulation with our proposed losses in regard to PSNR and perceptual metrics LPIPS and FID. Even our proposed single supervision loss in Fourier space $\mathcal{L_F}$, from our ablation study, substantially outperforms WaveletSRNet in three metrics with the exception of LPIPS, see configuration 2 in Tab.~\ref{tab:ablation}.




\begin{table}
\begin{center}
\setlength\tabcolsep{1.5pt}
\begin{tabular}{lcccc}
\toprule
Method & $\uparrow$PSNR & $\uparrow$SSIM & $\downarrow$LPIPS & $\downarrow$FID \\
\midrule
ESRGAN (Our losses) \cite{wang2018esrgan} & \textcolor{red}{25.05} & \textcolor{red}{0.738} & \textcolor{red}{0.120} & \textcolor{red}{24.07} \\

ESRGAN \cite{wang2018esrgan} & 24.36 & 0.717 & \textcolor{blue}{0.123} & \textcolor{blue}{25.50} \\
RankSRGAN (NIQE) \cite{Zhang_2019_ICCV} & 24.52 & 0.715 & 0.143 & 27.47 \\
Ours (Full) & \textcolor{blue}{24.69} & \textcolor{blue}{0.723} & 0.132 & 26.70 \\
\bottomrule
\end{tabular}
\end{center}
\caption{Evaluation on Urban100. \textcolor{red}{Red} indicates best, \textcolor{blue}{blue} second best.}
\label{tab:urban}
\end{table}


\begin{table*}
\begin{center}
\begin{tabular}{lccccrrr}
\toprule
Method & $\uparrow$PSNR & $\uparrow$SSIM & $\downarrow$LPIPS & $\downarrow$FID & $\downarrow$Par [M] & $\downarrow$GPU [s] & $\downarrow$CPU [s] \\
\midrule
Bicubic &28.11 & 0.782 & 0.410 &44.79 &-&-& -\\
SRFlow \cite{lugmayr2020srflow} & \textcolor{blue}{28.68} & 0.773 & 0.120 & 16.13 & 39.542  & 1.995 & 55.33\\
ESRGAN \cite{wang2018esrgan} & 28.19 & 0.769 & \textcolor{red}{0.115} & \textcolor{red}{15.37} & 16.698 & 0.553 & 29.28\\
RankSRGAN (Ma) \cite{Zhang_2019_ICCV} & 27.30 & 0.742 & 0.141 & 18.40 & \textcolor{blue}{1.554}  &\textcolor{blue}{0.099}& \textcolor{blue}{3.97}\\
RankSRGAN (NIQE) \cite{Zhang_2019_ICCV} & 28.19 & 0.765 & \textcolor{blue}{0.119} & 15.89 & \textcolor{blue}{1.554} &  \textcolor{blue}{0.099} & \textcolor{blue}{3.97}\\
RankSRGAN (PI) \cite{Zhang_2019_ICCV} & 28.11 & 0.765 & 0.121 & 16.28 & \textcolor{blue}{1.554} & \textcolor{blue}{0.099} & \textcolor{blue}{3.97}\\
\midrule
Ours (WaveletSRNet loss \cite{huang2017wavelet}) & 27.97 & \textcolor{blue}{0.786} & 0.171 & 19.80 & \textcolor{red}{0.894} & \textcolor{red}{0.041} & \textcolor{red}{1.72}\\
Ours (L1 only) & \textcolor{red}{30.56} & \textcolor{red}{0.837} & 0.270 & 22.91 & \textcolor{red}{0.894} &  \textcolor{red}{0.041} & \textcolor{red}{1.72} \\
Ours (Full) & 28.28 & 0.770 & 0.121 & \textcolor{blue}{15.72} & \textcolor{red}{0.894}  & \textcolor{red}{0.041} & \textcolor{red}{1.72} \\
\bottomrule
\end{tabular}
\end{center}
\caption{Comparison with state-of-the-art methods.  We compare in terms of image quality scores (PSNR, SSIM, LPIPS and FID) and efficiency measures (parameters and runtimes).  \textcolor{red}{Red} indicates best, \textcolor{blue}{blue} second best.}
\label{tab:results}
\end{table*}

We evaluate our losses on Urban100~\cite{Huang-CVPR-2015_urban100} in Tab.~\ref{tab:urban} to show the generalization capability of our approach. Our efficient setting achieves comparable performance also on this dataset. The application of our losses to ESRGAN again results in clear improvements in all 4 metrics by a substantial margin, especially in the restoration metrics.

\section{Conclusion}
We present two Fourier domain losses -- a supervision and a GAN loss -- to strengthen the training signal for the task of perceptual image SR. Our ablation study clearly shows the provision of complementary information during training in addition to the losses in spatial-domain. Due to the improved guidance, it is possible to train a significantly lower complexity -- and therefore faster -- network to achieve comparable performance of much larger networks, which we regard as an important property for many practical applications. The runtime of the generator backbone can be cut down to only 41ms, which is over $13\times$ faster than ESRGAN and $48\times$ faster than SRFlow on GPU. The clear separation of images into LF (retained) and HF (missing) content and therefore the direct emphasis on the missing high frequencies in Fourier space, imposed by our proposed losses, helps the SR network to generate plausible HF content. At the same time, we also apply the corresponding spatial losses to leverage the complementary local information, which results in even better perceptual quality.
To the best of our knowledge, we are the first to successfully apply a GAN-based loss directly in Fourier space for SR. We are convinced that further research into architectural improvements of our Fourier-space GAN-network can even further boost the effectiveness of our approach.


{\small
\bibliographystyle{ieee_fullname}
\bibliography{egbib}
}

\newpage
\appendix

\section*{\Large Supplementary Material}
We provide additional evaluations and visual results for our main models, that support the conclusions in the paper. 
In order to show the generalization capabilities of our approach, we provide results on a third dataset (BSD100).
Also, in addition to PSNR, SSIM, LPIPS and FID, we calculate popular no-reference metrics (Ma, NIQE, PI) on DIV2K and discuss their limitations for perceptual quality assessment. 
We also show an ablation study for our proposed Fourier space GAN architecture.

\section{Quantitative Evaluations}

The evaluation on 3 different datasets show the clear benefits and generalization capabilities of our proposed losses in comparison to previous approaches. The application of our losses directly in Fourier domain clearly improves not only the perceptual quality, but also the restoration quality at the same time.

\subsection{Urban100}

In Tab.~\ref{tab:urban} we show the results of additional methods on Urban100~\cite{Huang-CVPR-2015_urban100}. As already discussed, our losses show similar performance as on DIV2K (validation), which shows the generalizability of our proposed loss functions. Again, ESRGAN (Our losses) achieves the highest perceptual scores with a substantial improvement in FID of $-1.43$ over the version without our losses.
Our losses in conjunction with IMDN~\cite{Hui-IMDN-2019} achieve comparable results with SRFlow despite the enormous difference in runtime (41ms vs. 1995ms). 
Ours (Full), our efficient implementation, outperforms all versions of RankSRGAN in every metric, and is also faster at inference time.

\begin{table}[b]
\begin{center}
\setlength\tabcolsep{1.5pt}
\begin{tabular}{lcccc}
\toprule
Method & $\uparrow$PSNR & $\uparrow$SSIM & $\downarrow$LPIPS & $\downarrow$FID \\
\midrule
SRFlow \cite{lugmayr2020srflow} & \textcolor{red}{25.25} & \textcolor{blue}{0.735} & 0.127& 26.22\\
ESRGAN (Our losses) \cite{wang2018esrgan} & \textcolor{blue}{25.05} & \textcolor{red}{0.738} & \textcolor{red}{0.120} & \textcolor{red}{24.07} \\
ESRGAN \cite{wang2018esrgan} & 24.36 & 0.717 & \textcolor{blue}{0.123} & \textcolor{blue}{25.50} \\
RankSRGAN (Ma) \cite{Zhang_2019_ICCV} & 24.12  & 0.704 & 0.143 & 27.72\\
RankSRGAN (NIQE) \cite{Zhang_2019_ICCV} & 24.52 & 0.715 & 0.143 & 27.47 \\
RankSRGAN (PI) \cite{Zhang_2019_ICCV} & 24.47 & 0.716 & 0.139 & 27.84\\
Ours (Full) & 24.69 & 0.723 & 0.132 & 26.70 \\
\bottomrule
\end{tabular}
\end{center}
\caption{Evaluation on Urban100~\cite{Huang-CVPR-2015_urban100}. \textcolor{red}{Red} indicates best, \textcolor{blue}{blue} second best.}
\label{tab:urban}
\end{table}

\subsection{BSD100}

In addition to DIV2K(val) and Urban100, we evaluate the performance also on BSD100~\cite{bsd100}, another commonly used dataset, in Tab.~\ref{tab:bsd}. Again, ESRGAN (Our losses) performs best in terms of perceptual quality and also achieves high restoration quality. SRFlow has high restoration quality but can not compete in perceptual quality in comparison to all other methods.
Ours (Full) outperforms all RankSRGAN models in all metrics. Only RankSRGAN (NIQE) achieves a lower FID score. 

\begin{table}
\begin{center}
\setlength\tabcolsep{1.5pt}
\begin{tabular}{lcccc}
\toprule
Method & $\uparrow$PSNR & $\uparrow$SSIM & $\downarrow$LPIPS & $\downarrow$FID \\
\midrule
SRFlow \cite{lugmayr2020srflow} & \textcolor{red}{26.08} & \textcolor{red}{0.667} & 0.183 &  66.24 \\
ESRGAN (Our losses) \cite{wang2018esrgan} & \textcolor{blue}{25.79} & \textcolor{blue}{0.658} & \textcolor{red}{0.158} & \textcolor{red}{57.90}\\
ESRGAN \cite{wang2018esrgan} & 25.34 & 0.643 & \textcolor{blue}{0.161} & \textcolor{blue}{60.42}\\
RankSRGAN (Ma) \cite{Zhang_2019_ICCV} & 25.06 & 0.633 & 0.183 & 65.75 \\
RankSRGAN (NIQE) \cite{Zhang_2019_ICCV} & 25.52 & 0.642 & 0.178 & 61.52\\
RankSRGAN (PI) \cite{Zhang_2019_ICCV} & 25.48 & 0.643 & 0.175 & 63.97 \\
Ours (Full) & 25.66 & 0.656 & 0.172 & 62.25\\
\bottomrule
\end{tabular}
\end{center}
\caption{Evaluation on BSD100~\cite{bsd100}. \textcolor{red}{Red} indicates best, \textcolor{blue}{blue} second best.}
\vspace{-0.2cm}
\label{tab:bsd}
\end{table}

\subsection{DIV2K - No-reference Metrics}
No-reference metrics are handcrafted quality assessment tools, which quantify the image quality without comparison to a ground truth. However, these metrics are limited for objective image quality quantification for image super-resolution, because of the lacking reference. We would like to learn the true target distribution, which involves much more than learning for beautiful images. We therefore chose FID~\cite{Heusel2017fid} as a quality measure for distributional similarity which together with LPIPS quantifies perceptual quality.

We list the results for no-reference quality metrics Ma~\cite{Ma_Metric_2017}, NIQE~\cite{NIQE} and PI~\cite{Blau_2018_ECCV_Workshops} for DIV2K in Tab.~\ref{tab:noref}. As expected the RankSRGAN~\cite{Zhang_2019_ICCV} models perform the best, as they are explicitly trained for these metrics. Therefore, a direct comparison to all other methods is not fair. Interestingly, RankSRGAN (Ma) outperforms all other RankSRGAN models, even those that are trained for these specific metrics, which is unexpected. It is unclear to us why these inconsistencies arise, since RankSRGAN is trained on DIV2K for exactly these metrics.

Among the methods that are not explicitly trained for these metrics, our losses applied to ESRGAN and IMDN achieve the best results overall. Ours (Full) achieves the highest Ma and PI scores, ESRGAN (Our losses) achieves the best NIQE score.

\begin{table}
\begin{center}
\setlength\tabcolsep{1.5pt}
\begin{tabular}{lccc}
\toprule
Method & $\uparrow$Ma & $\downarrow$NIQE & $\downarrow$PI\\
\midrule
RankSRGAN (Ma) \cite{Zhang_2019_ICCV} & \textcolor{red}{6.8142} & \textcolor{red}{2.6143} & \textcolor{red}{2.9000}\\
RankSRGAN (NIQE) \cite{Zhang_2019_ICCV} & \textcolor{blue}{6.6923}
 & 2.7121 & 3.0099\\
RankSRGAN (PI) \cite{Zhang_2019_ICCV} & 6.6794 & \textcolor{blue}{2.6851} & \textcolor{blue}{3.0029}\\
\hline
SRFlow \cite{lugmayr2020srflow} & 6.5230 & 3.5421 & 3.5095\\
ESRGAN (Our losses) \cite{wang2018esrgan} & 6.5580  & 3.0388 & 3.2404\\
ESRGAN \cite{wang2018esrgan} & 6.5937 & 3.0918 & 3.2491 \\
Ours (WaveletSRNet losses)~\cite{huang2017wavelet} & 5.9682 & 4.9011 & 4.4664  \\
Ours (Full) & 6.6792 & 3.0836 & 3.2022\\
\bottomrule
\end{tabular}
\end{center}
\caption{Evaluation of no-reference metrics on DIV2K. \textcolor{red}{Red} indicates best, \textcolor{blue}{blue} second best.}
\vspace{-0.2cm}
\label{tab:noref}
\end{table}

\section{Visual Results}

We show a series of visual examples on all 3 datasets to asses the quality by visual inspection. In addition we provide PSNR and LPIPS for each method as quantitative metrics for restoration and perceptual quality respectively.

The metrics are in line with our quantitative evaluation overall. Note, we deliberately show some cases where the individual scores do not exactly match our overall quantitative evaluation. These differences on individual images arise due to variance among different strengths and weaknesses of each method.

The application of our losses in general improves the restoration- and perceptual quality, as can be seen by the examples of ESRGAN, ESRGAN (Our losses) and Ours (Full). Even our efficient setting with IMDN as generator achieves comparable performance to the larger model ESRGAN and especially the largest model SRFlow with highly improved runtimes. Ours (Full) in general also improves perceptual and restoration quality in comparison with RankSRGAN.
Additionally, we observed that SRFlow tends to generate noisy output, even in areas of uniform color.

\begin{figure*}[t]
\begin{center}
   \includegraphics[width=1\linewidth]{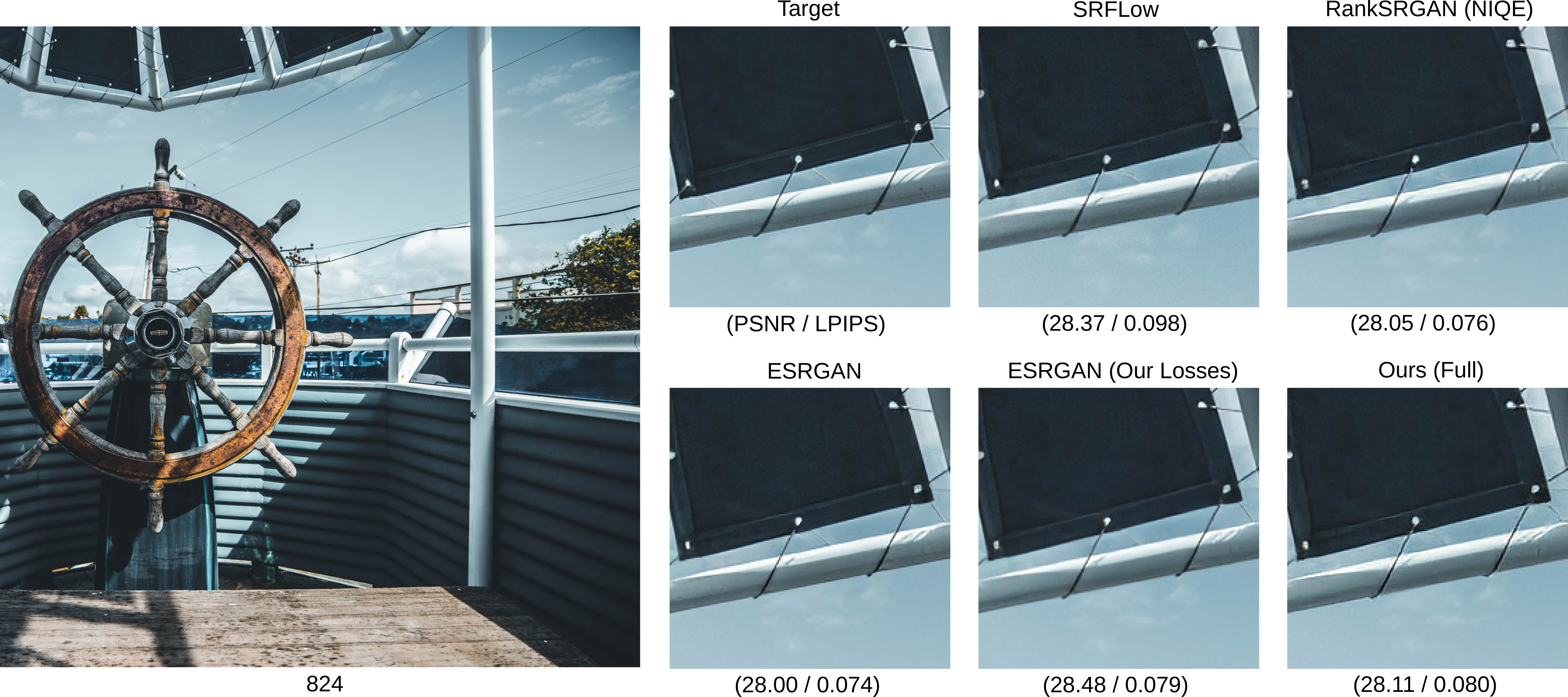}
\end{center}
   \caption{Visual examples on DIV2K, image 824.}
\label{fig:div824}
\end{figure*}

\begin{figure*}[t]
\begin{center}
   \includegraphics[width=1\linewidth]{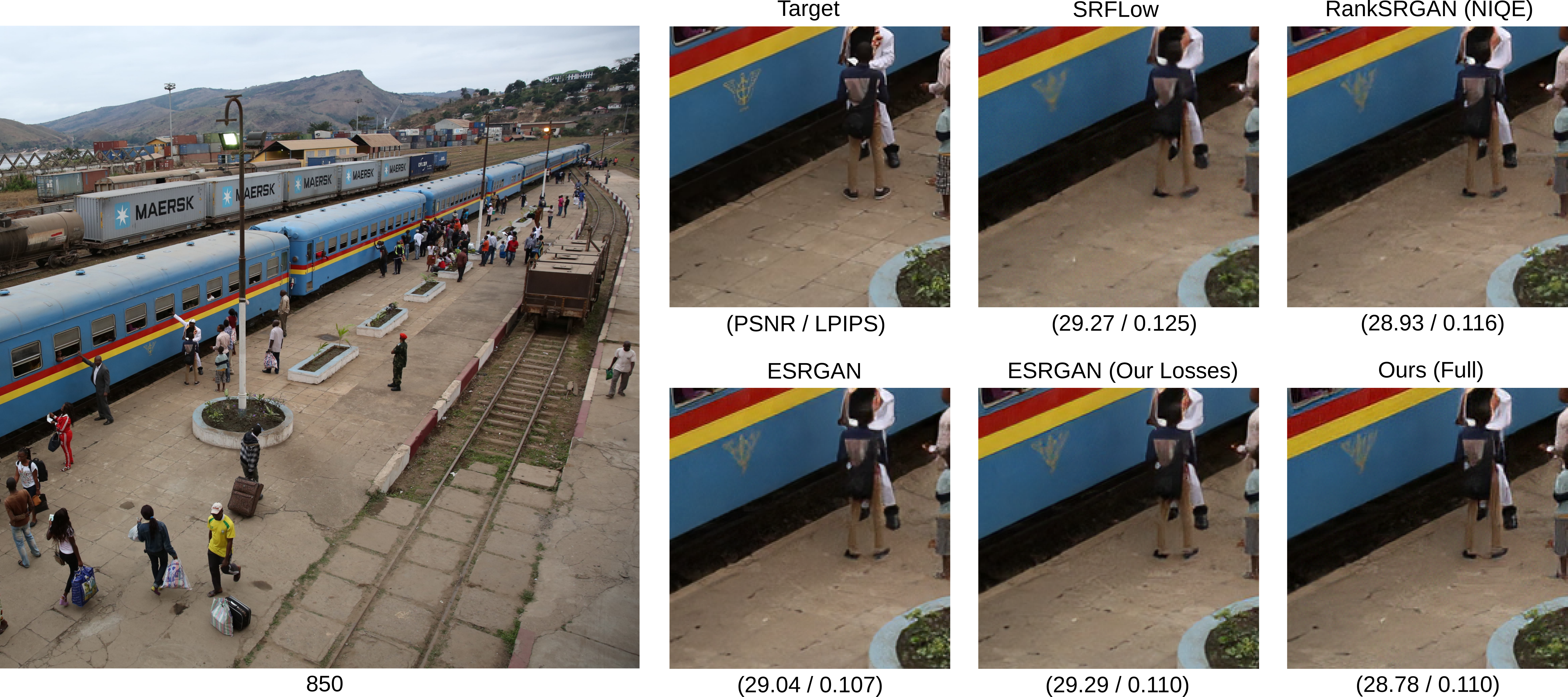}
\end{center}
   \caption{Visual examples on DIV2K, image 850.}
\label{fig:div850}
\end{figure*}

\begin{figure*}[t]
\begin{center}
   \includegraphics[width=1\linewidth]{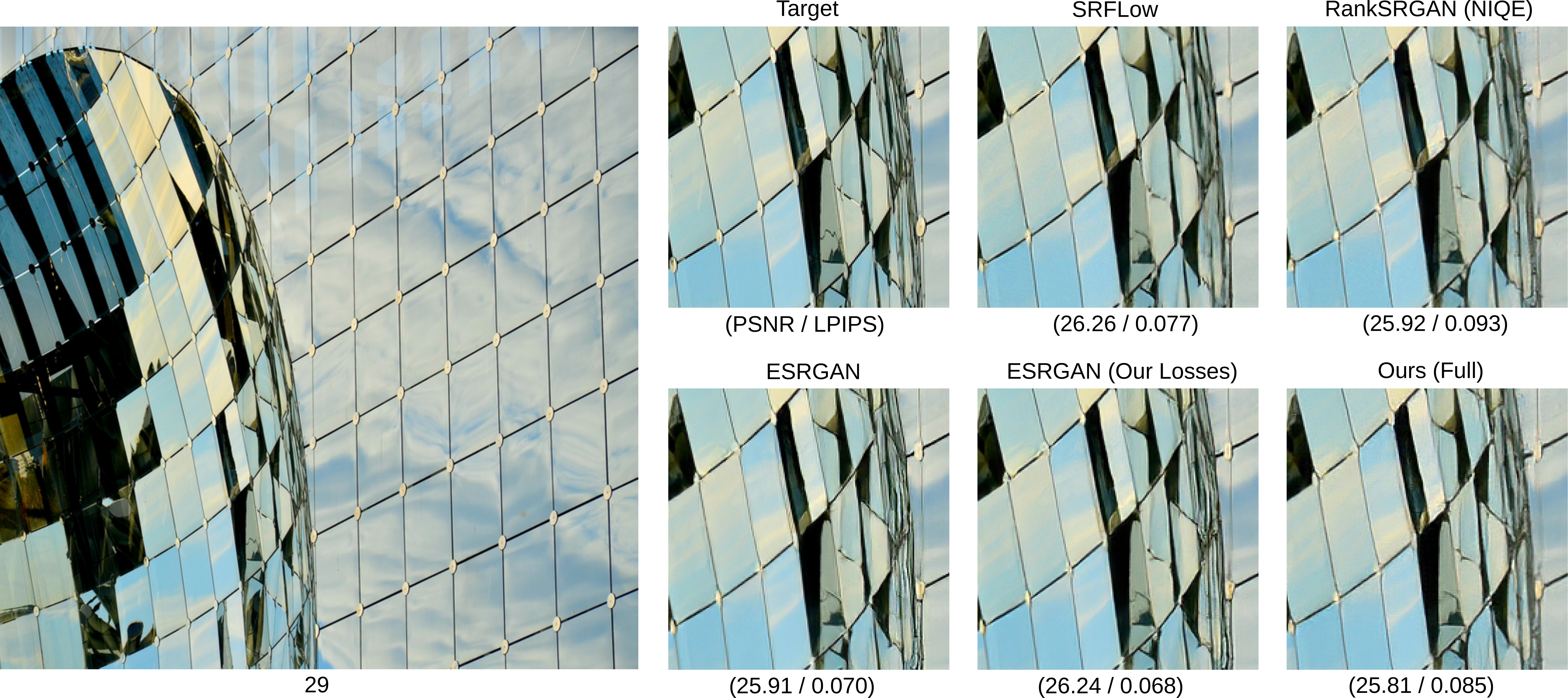}
\end{center}
   \caption{Visual examples on Urban100, image 29.}
\label{fig:urban29}
\end{figure*}

\begin{figure*}[t]
\begin{center}
   \includegraphics[width=1\linewidth]{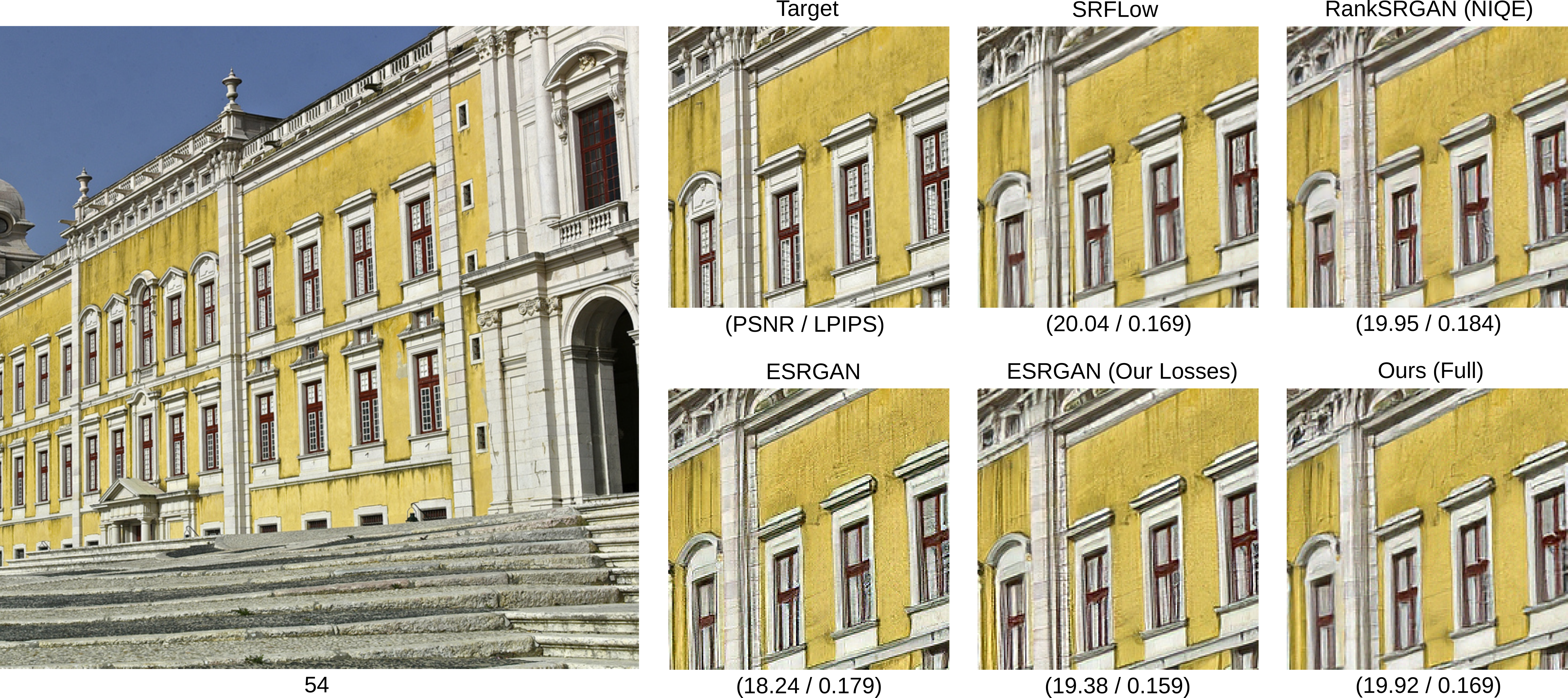}
\end{center}
   \caption{Visual examples on Urban100, image 54.}
\label{fig:urban54}
\end{figure*}

\begin{figure*}[t]
\begin{center}
   \includegraphics[width=1\linewidth]{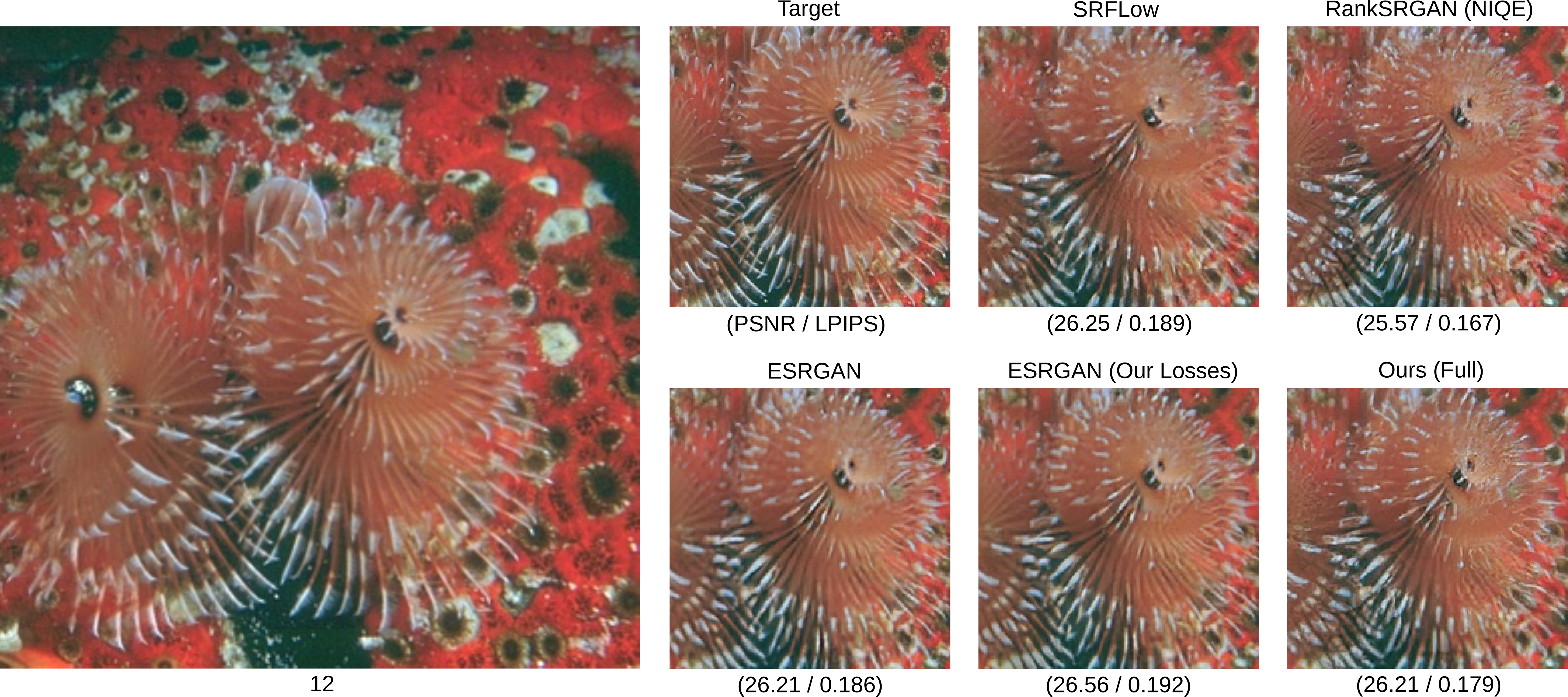}
\end{center}
   \caption{Visual examples on BSD100, image 12.}
\label{fig:bsd12}
\end{figure*}

\begin{figure*}[t]
\begin{center}
   \includegraphics[width=1\linewidth]{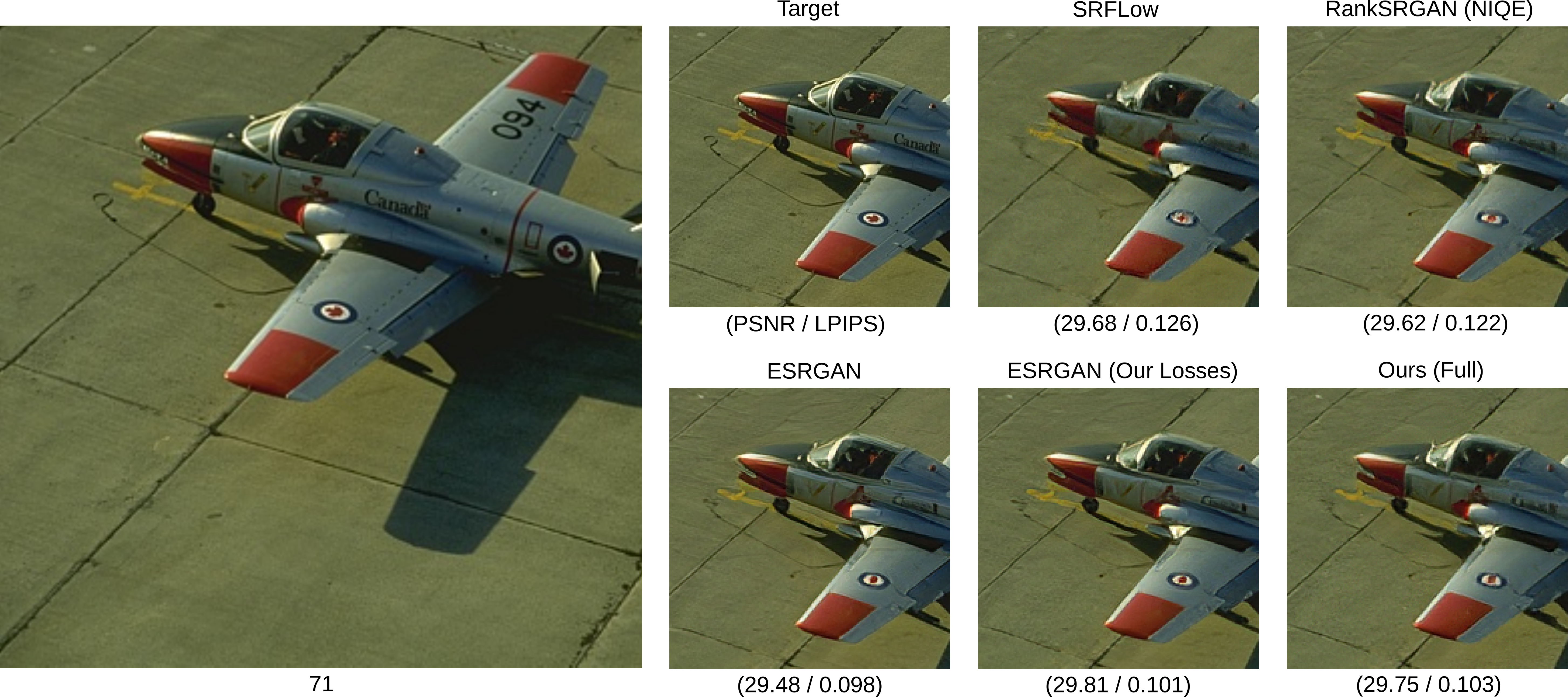}
\end{center}
   \caption{Visual examples on BSD100, image 71.}
\label{fig:bsd71}
\end{figure*}

\section{Fourier GAN Architecture - Ablation}

We provide additional analysis of our Fourier space GAN loss by training a smaller architecture with a reduced number of layers. We compare the full size GAN architecture ($\mathcal{L}^{\mathcal{F}, 5}_{GAN}$) with a reduced GAN architecture where the number of layers is set to 3 ($\mathcal{L}^{\mathcal{F}, 3}_{GAN}$).
We test this setup in configuration 5 and 8 from our ablation study in Tab.~\ref{tab:ablation}. The higher complexity discriminator achieves consistently better scores in PSNR, SSIM and FID in both configuration 5 and 8. LPIPS is slightly improved when using  $\mathcal{L}^{\mathcal{F}, 3}_{GAN}$. We suspect this could be in trade-off with FID due to an increased weight on the VGG-loss during training, when the discriminator is weaker.

\begin{table}[b]
\begin{center}
\setlength\tabcolsep{1.5pt}
\begin{tabular}{lcccc}
\toprule
Method & $\uparrow$PSNR & $\uparrow$SSIM & $\downarrow$LPIPS & $\downarrow$FID \\
\midrule

Ours (Config. 5, $\mathcal{L}^{\mathcal{F}, 3}_{GAN}$)  & 29.02 & 0.792 & \textcolor{red}{0.126} & 17.51 \\
Ours (Config. 5, $\mathcal{L}^{\mathcal{F}, 5}_{GAN}$)  & \textcolor{red}{29.06} & \textcolor{red}{0.796} &0.129& \textcolor{red}{17.17} \\

\midrule
Ours (Config. 8, $\mathcal{L}^{\mathcal{F}, 3}_{GAN}$)  & 28.32 & 0.770 & \textcolor{red}{0.122} & 16.19 \\
Ours (Config. 8, $\mathcal{L}^{\mathcal{F}, 5}_{GAN}$)  & \textcolor{red}{28.42} & \textcolor{red}{0.776} & 0.124 & \textcolor{red}{15.88} \\

\bottomrule
\end{tabular}
\end{center}
\caption{Ablation of FFTGAN architecture on  DIV2K~\cite{Agustsson_2017_CVPR_Workshops}. \textcolor{red}{Red} indicates best.}
\label{tab:ablation}
\end{table}

\end{document}